\documentclass[%
 reprint,
 superscriptaddress,
 amsmath,amssymb,
 aps,
 prl,
floatfix,
longbibliography 
]{revtex4-2}

\usepackage[pdftex]{graphicx} \graphicspath{{}}
\usepackage{float}
\usepackage{dcolumn}
\usepackage{bm}
\usepackage[pdftex,colorlinks=true]{hyperref}
\hypersetup{
  colorlinks=true,
  linkcolor=blue,
  urlcolor=cyan,
}

\newcommand{\eqn}[1]{\begin{equation} #1 \end{equation}}
\newcommand{\eqa}[1]{\begin{align} #1 \end{align}}

\usepackage{color}
\definecolor{ForestGreen}{RGB}{34, 139, 34}

\newcommand{\nn}{\nonumber}

\newcommand{\bS}{\boldsymbol{S}}

\newcommand{\bx}{\hat{\boldsymbol{x}}}

\newcommand{\bl}{\boldsymbol{l}}

\newcommand{\sectionn}[1]{\textit{#1---}}

\begin{document}

\preprint{APS/123-QED}

\title{
Domain wall dynamics in classical spin chains:\\ free propagation, subdiffusive spreading, and soliton emission}

\author{Adam J. McRoberts}
\affiliation{Max Planck Institute for the Physics of Complex Systems, N\"{o}thnitzer Str. 38, 01187 Dresden, Germany}

\author{Thomas Bilitewski}
 \affiliation{Department of Physics, Oklahoma State University, Stillwater, Oklahoma 74078, USA}
 
\author{Masudul Haque}
 \affiliation{Max Planck Institute for the Physics of Complex Systems, N\"{o}thnitzer Str. 38, 01187 Dresden, Germany}
 \affiliation{Institut f\"ur Theoretische Physik, Technische Universit\"at Dresden, 01062 Dresden, Germany}

\author{Roderich Moessner}
 \affiliation{Max Planck Institute for the Physics of Complex Systems, N\"{o}thnitzer Str. 38, 01187 Dresden, Germany}

 
\date{\today}

\begin{abstract}
\noindent
The non-equilibrium dynamics of domain wall initial states in a classical anisotropic Heisenberg chain exhibits a striking coexistence of apparently linear and non-linear behaviours:
the propagation and spreading of the domain wall can be captured quantitatively by {\it linear}, i.e.\ non-interacting, spin wave theory absent its usual justifications; while, simultaneously, for a wide range of easy-plane anisotropies, emission can take place of stable solitons---a process and objects intrinsically associated with interactions and non-linearities. 
The easy-axis domain wall only has transient dynamics, the isotropic one broadens diffusively, while the easy-plane one yields a pair of ballistically counter-propagating domain walls which, unusually,  broaden \textit{subdiffusively}, their width scaling as $t^{1/3}$. 
\end{abstract}

\maketitle

\begin{figure*}
    \centering
    \includegraphics{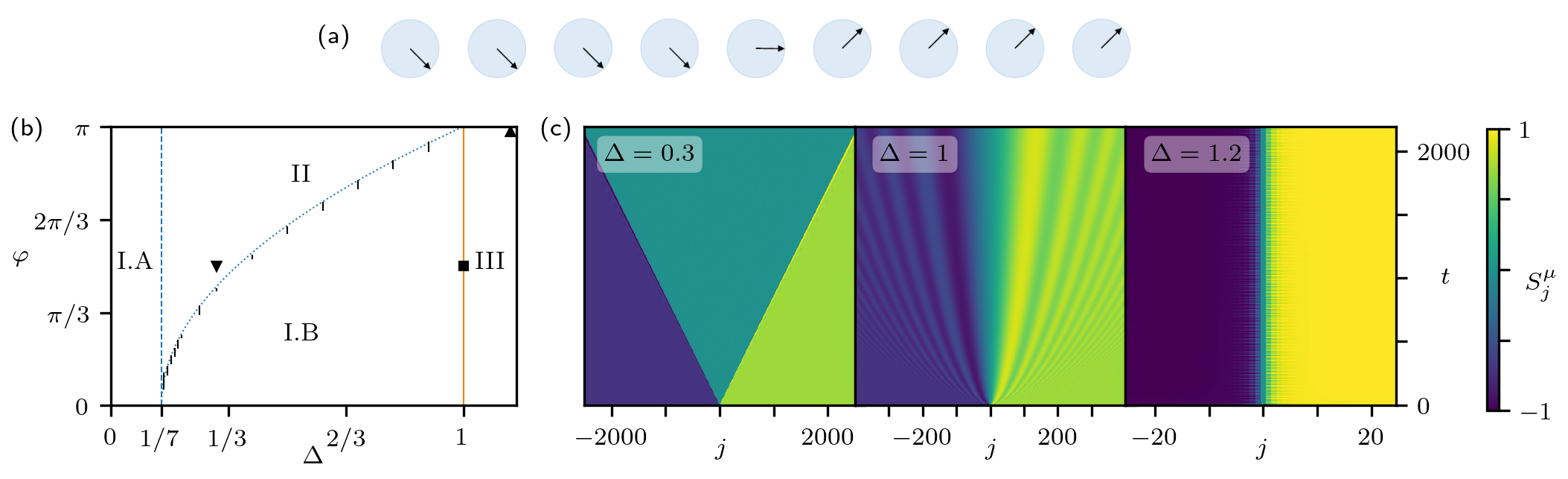}
    \caption{(a) Schematic of the initial conditions (\ref{eq:init_cond}), in the xy-plane, shown for $\varphi = \pi/2$. (b) Boundaries between the different dynamical regimes as a function of the anisotropy $\Delta$ and amplitude $\varphi$. I.A and I.B are the two linear regimes (distinguished by whether the oscillations are behind or ahead of the counter-propagating domain walls, respectively), where all of the dynamical features are well-described by linear spin wave theory; II is the easy-plane non-linear regime, where solitons coexist with the spreading domain walls. III is the easy-axis regime, with a single, static domain wall. The isotropic point $\Delta = 1$ corresponds to the transition between I.B and III, and is well-described by linear spin wave theory with a single, diffusively broadening domain wall. The vertical bars denote the uncertainty in determining the transition between I.B and II from the simulations. (c) Overview of the domain wall dynamics shown for the easy-plane $\Delta = 0.3$ ($\blacktriangledown$), the isotropic point $\Delta = 1$ ($\blacksquare$), and the easy-axis $\Delta = 1.2$ ($\blacktriangle$), respectively. Note the different ranges of the x-axes. $\varphi = \pi/2$ for the easy-plane and the isotropic case, where $S^y$ is plotted. $S^z$ is plotted for the easy-axis case, with $\varphi = \pi$. Ballistic counter-propagation is observed in the easy-plane, diffusive melting of the original domain wall is seen at the isotropic point, whilst the easy-axis approaches a very narrow static soliton. Note that, since the subdiffusive domain wall spreading in the easy-plane is parametrically slower than the ballistic propagation, and the emitted solitons move only very slightly slower than the domain walls and so are not well-separated over the timescales shown, both of these are difficult to see on this overview plot -- they are seen more readily in Fig.~\ref{fig:easy-plane} instead.}
    \label{fig:overview}
\end{figure*}

\sectionn{Introduction}
A prototypical setting for non-equilibrium dynamics is an initial state with two neighbouring regions in different stationary states of the same Hamiltonian.  A single sharp domain wall between distinct stationary states can move and spread, carrying energy---and, possibly, other conserved quantities. 
This type of dynamics has been studied in many contexts, including the spin-$\frac{1}{2}$ quantum XXZ chain \cite{
Antal_Racz_Schuetz_PRE1999, RaczSasvari_PRE2004_fronts, Kollath_Schollwoeck_Schuetz_PRE2005, Antal_Krapivsky_Rakos_PRE2008, Lancaster_Mitra_PRE2010, MosselCaux_NJP2010, Santos_Mitra_PRE2011, Sabetta_Misquich_PRB2013, eisler2013full,eisler2013universality, Alba_HeidrichMeisner_PRB2014, Viti_Stephan_EPL2016, Lancaster_PRE2016_NESS, Bertini_Collura_DeNardis_Fagotti_PRL2016,  JMStphan_JStatMech2017_ReturnProb, Ljubotina_Znidaric_Prosen_JPA2017, dePaulaPereiraDrummond_PRA2017, VidmarRigol_PRX2017emergent, LjubotinaZnidaricProsen_NatComm2017,  MisguichMallickKrapivsky_PRB2017, 
ColluraDeLucaViti_PRB2018Analytic, misguich2019domain, Bulchandani_Karrasch_PRB2019, JMStephan_SciPostPhys2019, Prosen_KPZ_PRL_2019, Gruber_Eisler_PRB2019, Moriya_Sasamoto_JSM2019, Collura_DeLuca_Calabrese_Dubail_PRB2020, AlbaBertiniFagottiPiroliRuggiero_JSM2021_ghd, Bulchandani_Gopalakrishnan_Ilievski_JSM2021, JMStephan_JPA2022_exact,  ScopaCalabreseDubail_JSM2022_doubleDWmelting, ScopaKarevski_EPJST2023_DWmelting}, 
other quantum spin chains \cite{EislerEvertz_SciPostPhys2016_IsingDW, Kormos_SciPostPhys2017_TransFieldIsing, Eisler_Maislinger_PRB2018_XY, Medenjak_DeNardis_PRB2020_spin1,  Eisler_ScipostPhys2020_XYfront, Bulchandani_Gopalakrishnan_Ilievski_JSM2021}, 
quantum field theories \cite{Lancaster_Gull_Mitra_PRB2010, Langmann_Lebowitz_Matropietro_CommMathPhys2016,  Horvath_Sotiriadis_Kormos_Takacs_SciPost2022_sineGordon},
the continuum Landau-Lifshitz model of classical spin densities \cite{Bulchandani_Gopalakrishnan_Ilievski_JSM2021,gamayun2019domain}, two-dimensional quantum systems \cite{Gambassi_Lerose_Scardiccio_PRL2022_2Dinterfacemelting, Gambasssi_Lerose_Scardiccio_PRB2023_2Dinterface}, and the simple exclusion process \cite{Derrida_JSP2009_SSEP_step}.  

Here, we consider a {\it classical} one-dimensional anisotropic (XXZ) Heisenberg chain,
\eqn{
H = -J \sum_i \left(S^x_i S^x_{i+1} + S^y_i S^y_{i+1} + \Delta S^z_i S^z_{i+1}\right),
\label{eq:H_XXZ}
}
where the $\bS_i \in S^2$ are classical $O(3)$ vectors at sites $i$ of a chain, and we assume $J > 0$ is ferromagnetic.
In addition to serving as one of the foundational models of magnetism and many-body spin physics, its dynamical properties have recently been subject to renewed investigation from various perspectives, including quantum-to-classical correspondence, non-linearity and integrability, and anomalous hydrodynamics \cite{Constantoudis_Theodorakopoulos_PRE1997,  Oganesyan_Pal_Huse_PRB2009_energytransport, Schmidt_Luban_JPCM2011_quasisolitons, Steinigeweg_EPL2012_spintransport, DeWijn_Hess_Fine_PRL2012_LLE, deWijn_Hess_Fine_JPA2013, Dhar_Huse_Moessner_Bhattacharjee_PRL2018, Damle_Dhar_Huse_JStatPhys2019, Bilitewski_Bhattacharjee_Moessner_PRB2021,  DeRaedt_Steinigeweg_PRB2021_QuantumVersusClassical, McRoberts_Bilitewski_PRB2022_anomalous, DeRaedt_Steinigeweg_PRR2022, McRoberts_Bilitewski_PRE2022_solitons}.

We investigate domain wall dynamics in this, arguably, simplest incarnation of this problem. We find a rich phenomenology with a number of intriguing aspects, and a co-existence of linear and non-linear behaviours: ballistically propagating domain walls which spread subdiffusively, showing the interplay of ballistic dynamics with subdiffusion, and which are well-described by \textit{linear} spin-wave theory; whilst, at the same time, we observe the emission of stable solitons, connecting to questions of dynamics and solitons in nearly-integrable systems.
It also opens a complementary perspective on the much studied related problem of quantum Heisenberg chains, where signatures of interesting phenomena such as KPZ scaling \cite{Kardar_PRL_1986,Prosen_KPZ_PRL_2019} have been experimentally observed \cite{KPZ_Bloch_Science_2022,Scheie_2021}. 

We find qualitatively distinct behaviour in the easy-plane, isotropic, and easy-axis cases ($0 \leq \Delta < 1$, $\Delta = 1$, and $\Delta > 1$, respectively).  Our results and set-up are summarised in Fig.~\ref{fig:overview}.
For easy-plane anisotropy, the domain wall splits into two ballistically counter-propagating ones (Fig.~\ref{fig:overview}(c)).  Since the Hamiltonian is non-integrable and intrinsically non-linear, and since the propagating domain walls have high energy compared to the background, they can, in principle, emit or decay into other excitations --- giving the non-equilibrium set-up an inherent non-linear flavour.   It is therefore all the more surprising that, over the entire range of easy-plane anisotropy $\Delta\in[0,1)$,  domain walls propagate ballistically.  This is reminiscent of the behaviour of quasiparticles in integrable systems \cite{ZotosPrelovsek_PRL1995, Calabrese_Cardy_JSTAT2005, Prosen_PRL_2011, Pereira_2014, Ilievski_PRL_2015, CastroAlvaredo_Doyon_PRX2016, Bertini_Collura_DeNardis_Fagotti_PRL2016,  Ilievski_2016, Alba_Calabrese_PNAS2017, Ilievaski_DeNardis_PRL2017,  Bulchandani_Gopalakrishnan_Ilievski_JSM2021, 
Bertini_FHM_Prosen_Steinigeweg_RMP2021, Gopalakrishnan_RPP_2023}, or that of operator spreading \cite{Gopalakrishnan_Huse_Khemani_Vasseur_PRB2018_HydroOperatorSpreading,Nahum_PRX2018_OperatorSpreading,vonKeyserlingk_Rakovsky_Pollmann_Sondhi_PRX2018}. For the latter, ballistic behaviour is accompanied by diffusive broadening \cite{Nahum_PRX2018_OperatorSpreading, vonKeyserlingk_Rakovsky_Pollmann_Sondhi_PRX2018, Gopalakrishnan_Huse_Khemani_Vasseur_PRB2018_HydroOperatorSpreading}. More generally, broadening in interacting many-body systems is typically diffusive, with exceptions usually associated with integrability or a lack of interactions.  

In sharp contrast to this expectation, we show that the  propagating domain walls broaden subdiffusively, as $\sim{t}^{1/3}$, in the entire easy-plane regime $\Delta\in[0,1)$.  
We find that the propagation speed, profile, and $t^{1/3}$ scaling can be quantitatively obtained from {\it linear} spin wave theory.  
At the same time, above a critical angle $\varphi_c(\Delta)$ between the domains separated by the propagating domain wall (Fig.~\ref{fig:overview}(b)), 
the linear behaviour of the propagating domain walls coexists with the aforementioned, inherently \textit{non}-linear feature of the emission of solitons. We provide a heuristic picture for all of these processes.

At the isotropic Heisenberg point $\Delta = 1$, the domain walls can no longer propagate, and the subdiffusive spreading gives way to a diffusive melting of the original domain wall (Fig.~\ref{fig:overview}(c)). Nor can the domain walls propagate in the easy-axis case ($\Delta > 1$), where the melting is fully arrested and a static soliton is approached asymptotically (Fig.~\ref{fig:overview}(c)).

The behaviour for $\Delta \geq 1$ is analogous to that known for quantum spin-$\frac{1}{2}$ chains \cite{Kollath_Schollwoeck_Schuetz_PRE2005, gamayun2019domain} -- a classical-quantum analogy which is, in itself, remarkable. By contrast, the $t^{1/3}$ broadening of the domain wall that we find over the entire range ${0\leq \Delta<1}$ appears, in the quantum spin-$\frac{1}{2}$ case, only at the ${\Delta=0}$ point \cite{RaczSasvari_PRE2004_fronts, eisler2013full, eisler2013universality,  Viti_Stephan_EPL2016} or at the light cone of fastest excitations \cite{Bulchandani_Karrasch_PRB2019, JMStephan_SciPostPhys2019, Bulchandani_Gopalakrishnan_Ilievski_JSM2021, Tracy_Widom_arxiv2202}, being associated with the  non-interacting (free-fermion) nature of these cases.  The existence and emission of the solitons have, to the best of our knowledge, not been previously observed -- either in the quantum model or in a corresponding continuum Landau-Lifshitz model.  

In the following, we provide details for these claims, and conclude with a discussion of the broader significance of this work.

\sectionn{Model}
We consider the classical XXZ spin chain, Eq.~\eqref{eq:H_XXZ}.
The dynamics is given by the classical equations of motion,
\eqn{
\dot{S}_i^{\mu} = -\epsilon^{\mu\nu\lambda} J_{\nu}(S^{\nu}_{i+1} + S^{\nu}_{i-1}) S_i^{\lambda},
\label{eq:EOM_XXZ}
}
which follow from the fundamental Poisson brackets $\{S^{\mu}_i, S^{\nu}_j\} = \delta_{ij}\epsilon^{\mu\nu\lambda}S^{\lambda}_j$, where $J_x = J_y = J = 1$ (which implicitly defines all units), and $0 \leq J_z = \Delta$.  The XY-point $\Delta = 0$ corresponds to the free-fermion limit of the quantum spin-$\frac{1}{2}$ chain, but is, in the classical case, an interacting model.

\sectionn{Easy-plane, $\Delta < 1$}
We consider a sharp domain wall in the in-plane components as the initial condition,
\eqa{
\bS_{i<0} &= \cos(\varphi/2)\hat{\boldsymbol{x}} - \sin(\varphi/2)\hat{\boldsymbol{y}} \nn \\
\bS_{i=0} &= \hat{\boldsymbol{x}} \nn \\
\bS_{i>0} &= \cos(\varphi/2)\hat{\boldsymbol{x}} + \sin(\varphi/2)\hat{\boldsymbol{y}}
\label{eq:init_cond}
}
for some amplitude $\varphi$ that sets the magnetisation jump across the domain wall as illustrated in Fig.~\ref{fig:overview}(a). The $O(2)$ isotropy implies that any choice of $\varphi$ connects two easy-plane ground states. We set the spin at $i = 0$ to lie halfway between the two domains, $\bS_{i=0} = \hat{\boldsymbol{x}}$ (though the results do not depend on the choice of $\bS_{i=0}$, so long as we do not select an unstable steady state~\footnote{Changing the direction of $\bS_{i=0}$ in the plane does not affect the results, since this only generates a small additional spin wave pulse, which rapidly dissipates; in particular, soliton formation is unaffected since the central domain is still $+\bx$, and soliton emission depends only on the angle between the central domain and the (+)- and (-)-domains.}).

Numerically integrating the equations of motion (\ref{eq:EOM_XXZ}) with initial conditions (\ref{eq:init_cond}) and open boundaries reveals that two counter-propagating domain walls immediately emerge from $i = 0$: a left-moving one connecting the $(-)$-domain to the expanding $\hat{\boldsymbol{x}}$-domain; and a right-moving one connecting the $\hat{\boldsymbol{x}}$-domain to the $(+)$-domain, as seen in Fig.~\ref{fig:overview}(c).

The size of the $\hat{\boldsymbol{x}}$-domain grows linearly with time, implying ballistic domain wall motion. Moreover, the domain-wall velocity does not differ measurably from the long-wavelength group velocity of the spin wave expansion, ${c = \sqrt{2(1 - \Delta)}}$ (cf. Eqs.~(\ref{eq:spin_wave_expansion}-\ref{eq:spin_wave_dispersion_parameters}), see also the Suppl. Mat.~\cite{supplemental}), despite the non-linearity of the equations of motion.

To investigate the long-time dynamics of the domain wall numerically, we switch to its co-moving frame \cite{supplemental}. We then find, numerically, that this easy-plane dynamics exhibits three qualitatively distinct regimes, cf. Fig.~\ref{fig:overview}(b): two linear regimes, I.A and I.B, so-called because they are well-described by \textit{linear} spin wave theory in their entirety; and a non-linear regime II characterised by an instability to the emission of solitons. 

\begin{figure}
    \centering
    \includegraphics{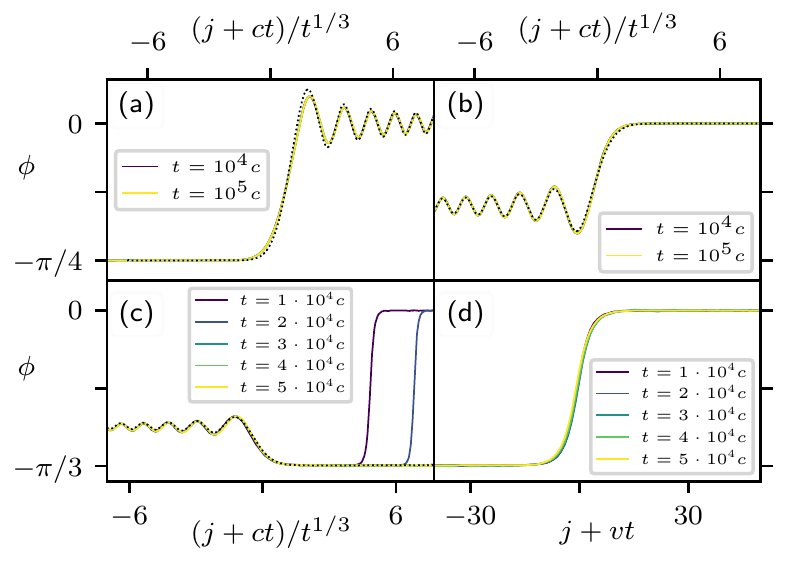}
    \caption{Dynamics in the easy-plane case. Dotted lines show spin wave predictions, where relevant. Only the left-moving domain walls are shown, and lighter colours indicate later times. (a) Linear regime I.A ($\Delta = 0$, $\varphi = \pi/2$), showing the ballistic propagation and subdiffusive spreading of the domain wall, with the oscillations trailing. (b) Linear regime I.B ($\Delta = 0.3$, $\varphi = 3\pi/10$), where the oscillations are now ahead of the domain wall. (c) Non-linear regime II ($\Delta = 0.25$, $\varphi = \pi/2$), showing that the domain wall decays by emitting a soliton, though its speed and width-scaling are unaffected; the soliton moves at a slower velocity $v < c$, and the separation increases with time. (d) Same parameters and times as (c), but in the co-moving frame of the soliton.}
    \label{fig:easy-plane}
\end{figure}

Within the linear regime we find, in addition to the ballistic motion of the domain walls, a sub-diffusive spreading, with their width scaling as $t^{1/3}$. We demonstrate this scaling collapse of the full domain wall profiles in Fig.~\ref{fig:easy-plane}(a,b). In the non-linear regime we observe the emission of a soliton during domain wall propagation shown in Fig.~\ref{fig:easy-plane}(c) which moves ballistically at a slower speed than the domain wall. We show with a purely ballistic scaling collapse in Fig.~\ref{fig:easy-plane}(d) that, indeed, this soliton does not disperse.

\sectionn{Spin-wave theory}
We next demonstrate that the spin-wave description of the easy-plane dynamics, remarkably, captures \textit{all} of the relevant features in what we call the linear regimes, and correctly predicts velocity and width-scaling of the domain walls even in the non-linear regime. 

We expand each spin about the $\bx$-domain,
\eqn{
\bS_i = \bx \sqrt{1 - l_i^2} + \bl_i,
\label{eq:spin_wave_expansion}
}
and retain only terms linear in $\bl_i$ in the equations of motion (which is, a priori, not controlled, as $\varphi$ is large!).

The analytical solution of the resulting problem is presented in the Suppl.\ Mat.\cite{supplemental}, but the central asymptotic result is readily stated: 
the spin-wave dispersion is given by
\eqn{
\omega_q \sim c|q| - \alpha|q|^3 + \dots, \;\;\; q \sim 0,
\label{eq:spin_wave_dispersion}
}
where 
\eqn{
c = 2\sqrt{1 - \Delta}, \;\;\; \alpha = \frac{1 - 7\Delta}{12c},
\label{eq:spin_wave_dispersion_parameters}
}
and, at long times, the left-moving domain wall is a function $\mathcal{D}$ of the variable $({j+c t})/{(3\alpha t)^{1/3}}$:
\eqn{
S_j^y(t) \sim \sin\left(\frac{\varphi}{2}\right) \mathcal{D}\left(\frac{j + c t}{(3\alpha t)^{1/3}}\right).
}
The linear spin wave calculation thus correctly predicts, asymptotically, two ballistically counter-propagating domain walls, each with a width scaling as ${w(t) \propto t^{1/3}}$. We also observe good quantitative agreement of the spin-wave prediction (dotted) with the profiles obtained in the full numerical simulation (solid curves) in Fig.~\ref{fig:easy-plane}(a,b).  The integral form is different from that appearing in the quantum free-fermion case \cite{Antal_Racz_Schuetz_PRE1999, Viti_Stephan_EPL2016} but is similar to those appearing in recent studies of caustics and catastrophes at light-cones \cite{Berry_2018, Kirkby_ODell_PRRes2019,  Farrell_Howls_ODell_JPA2023}.

\sectionn{Soliton emission in the non-linear regime}
\begin{figure}[t]
    \centering
    \includegraphics{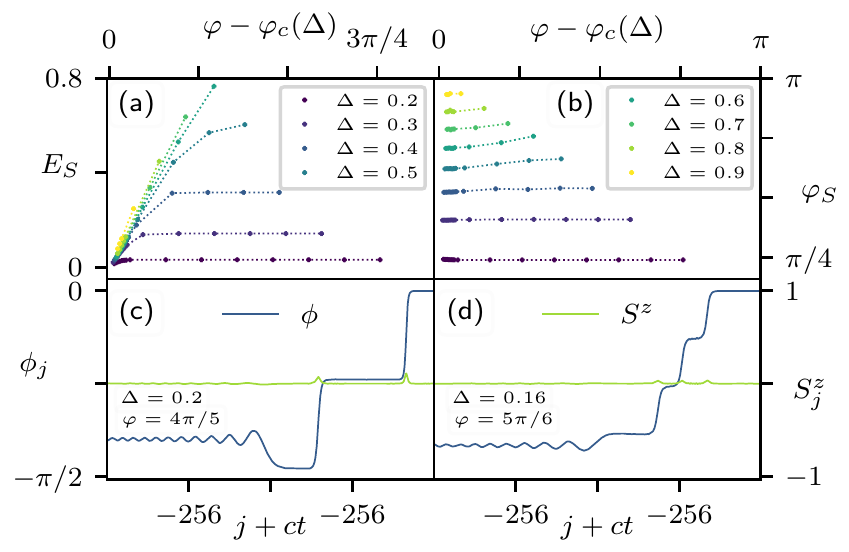}
    \caption{Soliton emission in the easy-plane dynamics. (a), (b) Dependence of the soliton energy $E_S$ and soliton amplitude $\varphi_S$, respectively, on the initial amplitude $\varphi$. We observe that the amplitude of the emitted solitons is almost constant. (c), (d) Two- and three-soliton emission, respectively, when ${\varphi \gg \varphi_S}$.}
    \label{fig:solitons}
\end{figure}
We next discuss the emission of solitons in the easy-plane regime. As observed in Fig.~\ref{fig:easy-plane}(c, d) the moving domain wall can emit a stable (non-dispersing) ballistically-propagating soliton connecting two ground states. We find that this emission only takes place above a critical, anisotropy-dependent amplitude $\varphi > \varphi_c(\Delta)$, and, in particular, only in the regime $1/7 < \Delta < 1$, as shown in Fig.~\ref{fig:overview}(b). The energy carried by these solitons is seen to depend both on the anisotropy $\Delta$ and the initial amplitude of the domain wall $\varphi$. Importantly, however, we observe that, at fixed $\Delta$, the angular amplitude $\varphi_S$ of the emitted soliton (the in-plane angle between the two ground states the soliton connects) does not depend on $\varphi$ (see Fig.~\ref{fig:solitons}(b)), and, in fact, is equal to the critical value $\varphi_c = \varphi_S$. Finally, we observe $n$-soliton emission if $\varphi > (2n-1)\varphi_S$, as shown in Figs.~\ref{fig:solitons}(c) \& \ref{fig:solitons}(d).

To explain this phenomenology we begin with an observation on the kinematics of magnons. When $\Delta < 1/7$, the spin-wave dispersion has \textit{negative} curvature at small $q$; this ensures that two-magnon scattering is elastic. In contrast, inelastic scattering is possible for $\Delta > 1/7$, allowing the dynamic instability towards soliton emission \cite{supplemental}.

To explain why the emitted soliton's amplitude $\varphi_S$ depends only on $\Delta$ (i.e., is unique for a given Hamiltonian), we propose the following heuristic model of soliton production. We assume that the model supports a two-parameter family of soliton solutions, which we may take to be their energy $E_S$ and velocity $v_S$. These two parameters, then, uniquely determine the other physical properties, such as the width and amplitude. Now, as the interactions are local, and the soliton is observed to be created at the ballistically-moving centre of the domain wall over an extended time, the speed of the soliton must initially be matched to the $\Delta$-dependent domain wall speed so that energy can be efficiently transferred---that is, $v_S = c(\Delta)$. Further, since the soliton is seeded by the domain wall, it must begin with zero energy $E_S \to 0$. This fixes the two parameters, and so picks out a unique initial soliton with some amplitude $\varphi_S(\Delta)$. As the dynamics proceeds, energy is transferred from the domain wall to the soliton, slowing down the latter and leading to its separation from the domain wall; but the amplitude $\varphi_S$ is a non-local property~\footnote{In this sense, that they connect distinct ground states, the solitons are topological---though we have not referred to them as such because they are not topologically \textit{protected}.}, and so cannot be changed by local dynamical processes after the soliton and domain wall begin to separate.

Finally, given the fixed soliton amplitude $\varphi_S$, we can provide an energetic argument for the stability regions. The domain wall energy depends monotonically on its amplitude, which must, therefore, decrease if soliton emission is to occur.  The initial amplitude of the domain wall is $\varphi/2$, and after the emission the new amplitude is $|\varphi_S - \varphi/2|$. Thus, emission is possible only if $\varphi > \varphi_c = \varphi_S$. This also implies that $n$-soliton emission is possible if $\varphi > (2n-1)\varphi_S$, as observed in Figs.~\ref{fig:solitons}(c) \& \ref{fig:solitons}(d).

\sectionn{Easy-axis and isotropic dynamics}
\begin{figure}[t]
    \centering
    \includegraphics{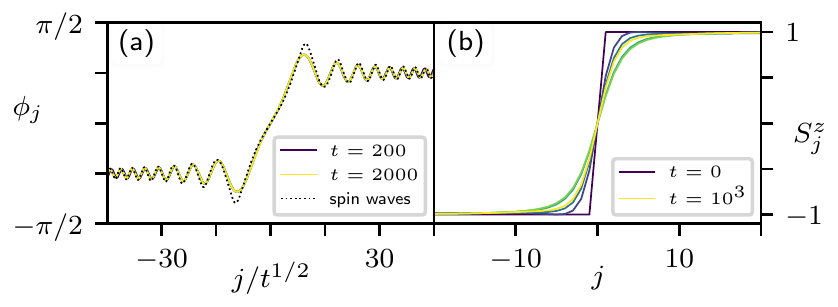}
    \caption{Dynamics (a) at the isotropic point, and (b) in the easy-axis regime ($\Delta = 1.2$). There are no propagating domain walls in either case: the initial state spreads diffusively at the isotropic point, whilst it approaches the static soliton in the easy-axis case.}
    \label{fig:isotropic}
\end{figure}
We briefly remark on the domain wall dynamics at the isotropic point (${\Delta = 1}$) and in the easy-axis case ({$\Delta > 1$}).

At the isotropic point, there can be no propagating ferromagnetic domain walls, because all components of the magnetisation are conserved; instead, we observe that the initially sharp domain wall spreads diffusively (Fig.~\ref{fig:isotropic}(a)). This can be understood within the linear spin-wave picture. At $\Delta = 1$, the dispersion switches from an odd-power expansion to the even expansion \cite{supplemental}
\eqn{
\omega_q = 2(1 - \cos(q)) \sim q^2 + ..., \;\;\; q \sim 0.
}  
There are no linear (dispersionless) terms---so the centre of the domain wall does not move---and the width is now controlled by the quadratic, not cubic, term.  Details of the calculation are presented in \cite{supplemental}.  The final, asymptotic answer can be conveniently expressed in terms of the normalised Fresnel integrals,
\eqn{
S_j^y(t) \sim \sin\left(\frac{\varphi}{2}\right) \left[\mathcal{C}\left(\frac{j}{\sqrt{2\pi t}}\right) + \mathcal{S}\left(\frac{j}{\sqrt{2\pi t}}\right) \right],
}
which shows good quantitative agreement with the full solution as seen in Fig.~\ref{fig:isotropic}(a).

For the easy-axis, we change the initial conditions so that the domain wall occurs in the $z$-components, ensuring that the state has finite energy. Specifically,
\eqn{
\bS_{i<0} = -\hat{\boldsymbol{z}}, \;\; \bS_{i=0} = \hat{\boldsymbol{x}}, \;\; \bS_{i>0} = +\hat{\boldsymbol{z}}.
}

Now, since the $z$-magnetisation is conserved, there can be no propagating domain wall solutions; some dissipative spin-wave radiation escapes, before the state settles down, in an oscillatory manner (Fig.~\ref{fig:isotropic}(b)), to the static soliton,
\eqn{
S_j^z = \tanh(j\cosh^{-1}(\Delta))
}
(see \cite{supplemental} for the derivation of this soliton solution).

\sectionn{Conclusions and outlook}
Our study of domain wall dynamics in the classical anisotropic Heisenberg spin chain reveals a remarkably rich and unexpected phenomenology, including ballistic propagation and sub-diffusive spreading of domain walls in the easy-plane regime, alongside the existence and emission of stable solitons---a highly non-linear phenomenon co-incident with a description of many aspects of the dynamics in the framework of \textit{linear} spin-wave theory.

The fact that all of the essential features of the regimes I.A and I.B (where no solitons are emitted, cf. Fig.~\ref{fig:overview}) are captured by a linearised description is, itself, remarkable---and connects this work to the broader question of under what conditions non-linear settings---e.g.\ \textit{a priori} beyond the linear response regime---may still be described by simplified linear theories.
This issue has appeared prominently, for example, in the study of KPZ dynamics \cite{KPZ_Bloch_Science_2022,Scheie_2021,Gopalakrishnan_RPP_2023} expected for small jumps in the initial condition, but in fact observed for larger ones. Further, how a description of the `doubly non-linear' phenomenon of the {\it emission} of (single or even multiple) {\it stable solitons} can co-exist with a linear description of the propagation of the emitting domain wall is a tantalising open question for future theoretical work.

This work also sheds some light on the question of when, and to what extent, classical treatments can account for \textit{a priori} complex quantum dynamics, by providing closely related instances of where this appears to be (im)possible: while the  $\Delta\geq1$ regimes and the XY point ($\Delta=0$) appear to be entirely analogous both classically and quantum-mechanically \cite{gamayun2019domain,misguich2019domain,ScopaCalabreseDubail_JSM2022_doubleDWmelting,JMStphan_JStatMech2017_ReturnProb,ScopaKarevski_EPJST2023_DWmelting}, the $0<\Delta<1$ regime is qualitatively distinct in the classical case. While reflecting some properties of the quantum $\Delta=0$ case \cite{eisler2013full, eisler2013universality}, the phenomenon of soliton emission has not been observed in previous studies of either the $S=\frac{1}{2}$ quantum case or the continuum Landau-Lifshitz model.  
   
Overall, it has become clear that spin chains, not just quantum but also classical, host many unexplored features.  The classical Heisenberg spin chain in particular has proven to be a fruitful platform to uncover and understand complex phenomena, and is presumably good for many surprises in future studies.  

\textcolor{blue}{}

\begin{acknowledgements}
This work was in part supported by the Deutsche Forschungsgemeinschaft under grants SFB 1143 (project-id 247310070) and the cluster of excellence ct.qmat (EXC 2147, project-id 390858490). This work was concluded at Aspen Center for Physics, which is supported by National Science Foundation grant PHY-2210452
\end{acknowledgements}

\nocite{Berry_2018,Farrell_Howls_ODell_JPA2023, Kirkby_ODell_PRRes2019}


\bibliography{refs} 

\onecolumngrid

  \cleardoublepage
  \begin{center}
    \textbf{\large Supplementary Material}
  \end{center}
\setcounter{equation}{0}
\setcounter{figure}{0}
\setcounter{table}{0}
\makeatletter
\renewcommand{\theequation}{S\arabic{equation}}
\renewcommand{\thefigure}{S\arabic{figure}}
\renewcommand{\thetable}{S\arabic{table}}
\setcounter{section}{0}
\renewcommand{\thesection}{S-\Roman{section}}

\setcounter{secnumdepth}{2}

\section*{Contents}

In the Supplementary Material, we describe in more detail the numerical procedure for switching to the co-moving frame of the ballistically propagating domain walls in the easy-plane dynamics; give an expanded account of the spin-wave calculations for the easy-plane and isotropic domain walls; expatiate on the kinematics of two-magnon scattering and the implications regarding soliton emission; and derive the exact expression for the easy-axis solitons.

\section{Co-moving frame}

In this section we describe the precise numerical procedure for switching to the co-moving frame of the (left-moving) domain wall in the easy-plane, which allows us to investigate the dynamics to much longer times than could be achieved on a fixed finite-size system.

We begin with the sharp domain wall initial conditions (3) on a finite chain of length $L$ (with sites $i \in [-\frac{L}{2}, \frac{L}{2}) \cap \mathbb{Z}$) and open boundaries. After a reset-time $t_R = L/4c$ (i.e., the time at which the left domain wall reaches site $i = -L/4$, where $c$ is the domain wall speed), we shift the state rightward by $L/4$, i.e., we map $\bS_i \mapsto \bS_{i-L/4}$, the rightmost quarter of the state is discarded, and the leftmost quarter is reset to the $(-)$-domain. We store a snapshot of this reset state, and then continue with the numerical time evolution, resetting periodically.

Clearly, resetting the state in this manner induces extra sources of error, viz., any information in the rightmost quarter is lost, and any information which reaches the finite left-boundary will be reflected. The (approximate) locality of the domain walls, however, ameliorates any potential problems here---since their width $w \sim t^{1/3}$ scales slowly with time, only modest system sizes $L \sim 10^4$ are needed to reach long times $t = 10^6$.

\section{Spin-wave calculation in the easy-plane case}

Let us now present the spin-wave calculation for the easy-plane domain wall dynamics in more detail. 

To ensure that the Fourier-modes are well-defined, we will work, as an intermediate step, with periodic boundary conditions for a system of size $L$, and initial conditions
\eqa{
\bS_{i=0} &= \hat{\boldsymbol{x}} \nn \\
\bS_{i=1,...,L/2-1} &= \cos(\varphi/2)\hat{\boldsymbol{x}} + \sin(\varphi/2)\hat{\boldsymbol{y}} \nn \\
\bS_{i=L/2} &= \hat{\boldsymbol{x}} \nn \\
\bS_{L/2+1,...L-1} &= \cos(\varphi/2)\hat{\boldsymbol{x}} - \sin(\varphi/2)\hat{\boldsymbol{y}}.
}
We will take the thermodynamic limit when it is explicitly safe to do so. Now, following the usual procedure of classical spin wave theory, we expand each spin about the $\bx$-domain,
\eqn{
\bS_i = \bx \sqrt{1 - l_i^2} + \bl_i,
}
and retain only terms linear in $\bl_i$ in the equations of motion (2). We obtain
\eqa{
\dot{l}_i^x &= 0, \nn \\
\dot{l}_i^y &= 2J_x l_i^z - J_z(l_{i+1}^z + l_{i-1}^z), \nn \\
\dot{l}_i^z &= -2J_x l_i^y + J_y(l_{i+1}^y + l_{i-1}^y).
}
These equations are readily solved by Fourier transformation. Since $l_q^z(0) = 0$, we have
\eqn{
l_q^y(t) = S_q^y(t) = \cos(\omega_q t)S_q^y(0),
}
where the dispersion is given by
\eqn{
\omega_q = 2\sqrt{(1 - \cos q)(1 - \Delta \cos q)}.
}

The procedure now is to Fourier transform the initial conditions, apply the time-evolution to each $q$-mode, and then invert the transform. After some algebra, we obtain
\eqn{
S_j^y(t) = \frac{-2i}{L}\sin\left(\frac{\varphi}{2}\right)\sum_{n=1,3,...}^{L-1} e^{2\pi i n j/L} \cot\left(\frac{n\pi}{L}\right)\cos(\omega_{q}t),
}
where $q = 2\pi n/L$. We may now safely pass to the thermodynamic limit $L \to \infty$, which yields

\eqn{
S_j^y(t) = -2i\sin\left(\frac{\varphi}{2}\right) \int_{-\pi}^{\pi}\; \frac{dq}{4\pi} e^{iqj}\cos(\omega_q t) \cot\frac{q}{2}.
\label{eq:spin_wave_int}
}

To make further analytic progress, we expand the dispersion in powers of $q$:
\eqn{
\omega_q \sim c|q| - \alpha|q|^3 + \gamma|q|^5 + ...,\;\;\; |q| \sim 0,
}
where the coefficients are given as functions of $\Delta$ by
\eqn{
c = \sqrt{2(1-\Delta)}, \;\;\; \alpha = \frac{1-7\Delta}{12 c}
,\;\;\; \gamma = \frac{1 - 62\Delta + \Delta^2}{960(1-\Delta)c}
.
}

Let us assume, for now, that $\alpha > 0$ (i.e., $0 < \Delta < 1/7$). We insert the expansion of the dispersion at third order, and note that the modulus signs can be ignored because the cosine is even. Then, splitting the cosine to make clear how this separates into left- and right-moving domain walls, we have
\eqn{
S_j^y(t) \sim -2i\sin\left(\frac{\varphi}{2}\right) \sum_{\sigma=\pm }\int_{-\pi}^{\pi}\; \frac{dq}{8\pi} e^{iq(j + \sigma c t) - i\sigma\alpha q^3 t} \cot\frac{q}{2}.
\label{eq:spin_wave_int_part}
}
We make the substitution
\eqn{
k^3 = 3\alpha q^3 t \;\;\Rightarrow\;\; q = \frac{k}{(3\alpha t)^{1/3}},
}
whereupon the integral becomes
\eqn{
S_j^y(t) \sim \sum_{\sigma=\pm}\int_{-(3\alpha t)^{1/3}\pi}^{(3\alpha t)^{1/3}\pi}\; \frac{dk}{4\pi} \exp\left(ik\frac{j + \sigma c t}{(3\alpha t)^{1/3}}\right) \left[\frac{-2i\sin\left(\frac{\varphi}{2}\right) e^{- i\sigma k^3/3}}{2(3\alpha t)^{1/3}}
\cot\left(\frac{k}{2(3\alpha t)^{1/3}}\right) \right].
\label{eq:spin_wave_sub_int}
}

We now consider the $t \to \infty$ asymptotics. We send the limits of the integrals to $\pm \infty$, and use the asymptotic equivalence
\eqn{
\frac{1}{X}\cot\left(\frac{k}{X}\right) \sim \frac{1}{k},\;\;\; X \to \infty,
\label{eq:cot_rel}
}
which yields
\eqn{
S_j^y(t) \sim \sin\left(\frac{\varphi}{2}\right) \sum_{\sigma=\pm} \mathcal{F}^{-1}_{[f_\sigma]}\left(\frac{j+\sigma c t}{(3\alpha t)^{1/3}}\right),
\label{eq:spin_wave_easy_plane}
}
where
\eqn{
\mathcal{F}^{-1}_{[f]}(X) := \int_{\mathbb{R}} \frac{dk}{2\pi}\;e^{ikX}f(k), \;\;\; f_\sigma(k) = \frac{-ie^{-i\sigma k^3 / 3}}{k}.
\label{eq:ifft}
}

We note that this integral has been previously considered in the context of undular tidal bores \cite{Berry_2018}, and more generally in the context of catastrophes in waves at horizons \cite{Farrell_Howls_ODell_JPA2023}, and related integrals generically describe light cones in quenches in quantum spin chains \cite{Farrell_Howls_ODell_JPA2023,Kirkby_ODell_PRRes2019}.

The calculation is the same for $\alpha < 0$ ($\Delta > 1/7$), except that an overall minus sign is attached to Eq.~\eqref{eq:spin_wave_easy_plane}, since we have to flip over the integration limits in Eq.~\eqref{eq:spin_wave_sub_int}. The functions $\mathcal{F}^{-1}_{[f_{\sigma}]}$ can be explicitly evaluated in terms of the generalised hypergeometric functions $_{p}F_{q}$, though the full expressions are somewhat unwieldy; it is apparent, however, that the spin-wave calculation correctly predicts two ballistically counter-propagating domain walls which broaden subdiffusively, and, as shown in Fig.~2 of the main text, reproduces the full profile reasonably well. The spin wave calculation, cannot, of course, predict the emission of solitons, since they are intrinsically non-linear objects.

Precisely at the point $\Delta = 1/7$, the cubic term vanishes and an analogous calculation would predict that the width should scale as $t^{1/5}$. We do not observe this in the numerical simulations (which use the full equations of motion), which may be due to non-linear interactions between the spin wave modes renormalising the dispersion.

\section{Spin-wave calculation at the isotropic point}

The spin-wave calculation at the isotropic point proceeds with exactly the same steps (initial conditions, equations of motion, Fourier transform, and thermodynamic limit) as the easy-plane case---up to the insertion of the dispersion relation,
\eqn{
\omega_q = 2(1 - \cos(q)) \sim q^2 + ..., \;\;\; q \sim 0,
}  
into Eq.~(\ref{eq:spin_wave_int}). From this point, we have
\eqn{
S_j^y(t) = -2i\sin\left(\frac{\varphi}{2}\right) \int_{-\pi}^{\pi}\; \frac{dq}{4\pi} e^{iqj}\cos(q^2 t) \cot\frac{q}{2},
\label{eq:spin_wave_int_iso}
}
and we now use the substitution 
\eqn{
k^2 = 2q^2 t, \;\; \Rightarrow \;\; q = \frac{k}{(2t)^{1/2}}.
}
Together with the asymptotic relation (\ref{eq:cot_rel}), this yields
\eqn{
S_j^y(t) = \sin\left(\frac{\varphi}{2}\right) \mathcal{F}^{-1}_{[g]}\left(\frac{j}{(2t)^{1/2}}\right),
}
in the notation of Eq.~(\ref{eq:ifft}), with
\eqn{
g(k) = \frac{-2i}{k}\cos\left(\frac{k^2}{2}\right);
}
or, in terms of the normalised Fresnel integrals,
\eqn{
S_j^y(t) = \sin\left(\frac{\varphi}{2}\right) \left[\mathcal{C}\left(\frac{j}{\sqrt{2\pi t}}\right) + \mathcal{S}\left(\frac{j}{\sqrt{2\pi t}}\right) \right].
}

\section{Magnon kinematics}

We turn now to some observations on the kinematics of two-magnon scattering, and the implications for soliton emission observed numerically. We consider two-magnon scattering, since these are the lowest-order processes which could lead to some dynamic instability.

For two magnons with momenta $q_1$, $q_2$, which scatter to two magnons with momenta $q_3$, $q_4$, conservation of momentum implies that we may write these as:
\eqa{
&q_1 = p \;\;\;\;\;\;\;\;\;\;\;\;\;\;q_2 = p' \nn \\
&q_3 = p' - q \;\;\;\;\;\;\;q_4 = p + q
}
for some momentum transfer $q$. On the other hand, energy conservation imposes a further constraint:
\eqn{
\omega_{p+q} + \omega_{p'-q} = \omega_{p'} + \omega_p.
\label{eq:energy_conservation}
}

Let us assume that all of the magnon momenta have the same sign, and, without further loss of generality, that they are all positive. For soliton formation we do not consider back-scattering, since any scattering processes must be local, and left and right-movers quickly separate in the sharp domain wall set-up we consider; they cannot, therefore, contribute to the dynamical formation of a soliton.

Now, we use the fifth-order expansion of $\omega_k$,
\eqn{
\omega_k \sim c k - \alpha k^3 + \gamma k^5.
\label{eq:fifth_order_dispersion}
}
Inserting this into Eq.~\eqref{eq:energy_conservation} yields a quartic equation for the allowed momentum transfers $q$ in terms of $p$, $p'$, $\alpha$, and $\gamma$ (the latter two being functions of $\Delta$). There are two trivial (elastic) solutions,
\eqn{
q = 0,\;\;\; q = p' - p,
}
and two non-trivial, but potentially complex solutions,
\eqn{
q = \frac{p'-p}{2} \pm \frac{\sqrt{5}\sqrt{12\alpha\gamma - 5\gamma^2\left(3p^2 + 2p p' + 3p'^2\right)}}{10\gamma}.
\label{eq:scattering_solutions}
}

Now, for $\Delta > 1/7$ ($\alpha < 0$, $\gamma < 0$) there is a range of momenta for which inelastic scattering is possible, given by the condition
\eqn{
\frac{12}{5}\frac{|\alpha|}{|\gamma|} > 3p^2 + 2p p' + 3p'^2 > \frac{12}{5}\frac{|\alpha|}{|\gamma|} - (p + p')^2.
\label{eq:scattering_region}
}
However, inelastic scattering is not possible for $\Delta < 1/7$ ($\alpha > 0$). Over most of this region, $\gamma < 0$, and the solutions \eqref{eq:scattering_solutions} are complex. There is a very small region near $\Delta = 0$ where $\gamma > 0$, but here the solutions given by Eq.~\eqref{eq:scattering_solutions} violate the assumption (needed to remove the modulus signs in Eq.~\eqref{eq:fifth_order_dispersion}) that all of the magnon momenta are positive (either $p + q < 0$ or $p' - q < 0$, depending on the branch -- this is also the origin of the second inequality in Eq.~\eqref{eq:scattering_region}).

We conjecture that the fact that two-magnon scattering is always elastic for $\Delta < 1/7$ explains why regime I.A (cf. Fig.~1(b)) is never observed to emit solitons, for any initial domain wall amplitude $\varphi$, since the main dynamical processes that would manifest such an instability are forbidden.

\section{Easy-axis soliton}

We close by providing the derivation of the exact easy-axis soliton (11) on the lattice, which has a subtly different expression to its continuum counterpart. 

We begin with the ansatz that all of the spins are static,
\eqn{
\dot{\bS}_j = 0, \;\;\; \forall j,
}
and, moreover, that all of the in-plane components point in the same direction, which, without loss of generality, we take to be $+\bx$. That is, we have
\eqn{
S_j^x(t) = \sqrt{1 - z_j^2}, \;\;\; S_j^y(t) = 0, \;\;\; S_j^z(t) = z_j.
}
Substituting this ansatz into the full equations of motion (2), we find that $\dot{S}_j^x = 0$ and $\dot{S}_j^z = 0$ $\forall j$ are automatically satisfied (since all of the $y$-components vanish by assumption). The requirement that the $y$-components, indeed, remain zero for all time then gives rise to a set of consistency equations for the $z_j$,
\eqn{
\dot{S}^y_j = 0 \;\;\; \Rightarrow \;\;\; \frac{z_j}{\sqrt{1 - z_j^2}} = \Delta \frac{z_{j+1} + z_{j-1}}{\sqrt{1 - z_{j+1}^2} + \sqrt{1 - z_{j-1}^2}},
\label{eq:con}
}
for which, choosing the centre to be $z_0 = 0$ and setting the boundary conditions $z_{\pm\infty} = \pm 1$, the exact solution is
\eqn{
z_j = \tanh(j \; \mathrm{arccosh}(\Delta)),
}
assuming $\Delta > 1$, which may be verified by direct substitution into the consistency equations \eqref{eq:con}.

\end{document}